# National social cost of carbon: An application of FUND

In Chang Hwang, Richard S.J. Tol

In Chang Hwang

The Seoul Institute, 57 Nambusunhwan-ro, 340-gil, Seocho-gu, Seoul, 06756, South Korea,

ichwang@si.re.kr

Richard S. J. Tol

Department of Economics, University of Sussex, Falmer, UK, r.tol@sussex.ac.uk

Institute for Environmental Studies, Vrije Universiteit, Amsterdam, The Netherlands

Department of Spatial Economics, Vrije Universiteit, Amsterdam, The Netherlands

Tinbergen Institute, Amsterdam, The Netherlands

CESifo, Munich, Germany

Payne Institute for Public Policy, Colorado School of Mines, Golden, CO, USA

## Abstract

This paper presents a refined country-level integrated assessment model, FUND 3.9n, that extends the regional FUND 3.9 framework by incorporating sector-specific climate impact functions and parametric uncertainty analysis for 198 individual countries. The model enables estimation of the national social cost of carbon (NSCC), capturing heterogeneity across nations from economic structure, climate sensitivity, and population exposure. Our results demonstrate that both the NSCC and the global sum estimates are highly sensitive to damage specifications and preference parameters, including the pure rate of time preference and relative risk aversion. Compared to aggregated single-sector approaches, the disaggregated model with uncertainty yields higher values of the NSCC for low- and middle-income countries. The paper contributes to the literature by quantifying how sector-specific vulnerabilities and stochastic variability amplify climate damages and reshape global equity in the distribution of the NSCC. The NSCCs derived from our model offer policy-relevant metrics for adaptation planning, mitigation target setting, and equitable burden-sharing in international climate negotiations. This approach bridges the gap between globally harmonized carbon pricing and nationally differentiated climate impacts, providing a theoretically grounded and empirically rich framework for future climate policy design.

## Key words

Social cost of carbon, Climate impacts, Country-level SCC, FUND model

# 1 Introduction

A globally harmonized carbon tax, based on an optimal balance of marginal costs and benefits, is a key yardstick for international environmental governance (a cooperative solution). However, with fractious geopolitics, countries may seek a more self-interested noncooperative solution based purely on their own costs and benefits. This would lead to heterogeneous carbon pricing that would accommodate the diverse economic and environmental landscapes of each country. While heterogeneous carbon pricing allows for tailored strategies, it introduces complex dynamics where emissions from individual countries likely increase, leading to aggregate emissions exceeding optimal levels. It is important to understand what carbon prices countries would choose in this noncooperative solution. This underscores the necessity to understand the pricing mechanisms that align with purely national climate objectives.

A central component in determining optimal carbon pricing is the social cost of carbon (SCC), which quantifies the economic, social, and environmental damages associated with an additional tonne of carbon emissions. The SCC serves as a foundational metric for internalizing the external costs of carbon emissions, guiding policy decisions, and assessing the economic feasibility of climate mitigation efforts (e.g., Pearce, 2003; Tol, 2008; 2011; 2023; Weitzman, 2014; Nordhaus, 2014; 2017; Pizer et al., 2014; Adler et al., 2017; Kotchen, 2018; Pindyck, 2019; Cai and Lonzek, 2019; Aldy et al., 2021; Lemoine, 2021; Rode et al., 2021; Rennert et al., 2022; Barrage and Nordhaus, 2024). The SCC is a crucial tool for formulating carbon regulations and adaptation policies (IWG, 2016; EPA, 2023). But the SCC is based on an optimal cooperative solution.

If the world does not follow a cooperative solution but instead turns to a noncooperative solution, it is critical to measure the national (or country-level) SCC (NSCC). The NSCC is the economic, social, and environmental damages associated with an additional tonne of carbon emissions to each country. There is a different NSCC for each country. The distribution of the NSCC across countries reflects the heterogenous impacts of climate change to economies and populations across countries (OECD, 2015; IPCC, 2023). In order to measure the outcome across countries, one must measure the NSCC for each country which requires more spatial detail than most economic IAM models have generated in the past.

An important exception is Ricke et al. (2018) and Tol (2019) who have already introduced national SCC estimates. These papers highlight the distributional aspects of climate impacts and incentives influencing national climate policies. However, they rely on a single, aggregate impact function, without sectoral specificity. Considering nonlinear sectoral functions (Tol, 2002a, b), the absence of sectoral uncertainty assessments hinders a comprehensive understanding of how variations in socioeconomic activities and climate projections shape national climate risks.

This paper addresses these gaps by developing a refined national SCC model that integrates sectoral impacts with uncertainty analysis. This approach acknowledges the inherent differences in socioeconomic situations, climate vulnerabilities, and adaptive capacities across countries, thereby yielding more detailed NSCC estimates. Furthermore, this paper highlights how a higher resolution in sectoral and geographical details influences the calculation of the global SCC. The distinction between aggregated NSCCs and the global SCC derived from a single-region model is formalized through the following equation. In addition, incorporating uncertainty leads to a different NSCC estimation from using best-guess values.

$$SCC_g^b = \sum_n NSCC_n^u = \sum_n f(\boldsymbol{v}_n) \neq f\left(\sum_n \boldsymbol{v}_n\right) = SCC_g$$

In addition, incorporating uncertainty leads to a different SCC estimation from using best-guess values:

$$SCC_n^u = \mathbb{E}_{\boldsymbol{s}} f(\boldsymbol{v}_n(\boldsymbol{s})) \neq f(\mathbb{E}_{\boldsymbol{s}} \boldsymbol{v}_n(\boldsymbol{s})) = SCC_n$$

where $SCC_n^u$ denotes the national SCC derived from an uncertainty model, $SCC_g^b$ is the global SCC derived from an aggregation of NSCCs, $SCC_g$ is the global SCC estimated from a global single region model, $f$ is a nonlinear SCC function, $\boldsymbol{v}_n$ is the vector of state variables, $\mathbb{E}_{\boldsymbol{s}}$ is the expectation operator capturing uncertainty in parameters $\boldsymbol{s}$.

To implement this framework, we modify the FUND (Framework for Uncertainty, Negotiation, and Distribution) model (Anthoff and Tol, 2014) into a country-specific structure that integrates sectoral impacts. Our approach enhances the understanding of the intricate dynamics linking economic growth, carbon emissions, and climate impacts at the national level. This improved granularity provides critical insights for policymakers aiming to devise more effective mitigation and adaptation strategies that align with national priorities while contributing to global climate objectives.

The paper is structured as follows. Section 2 details the model framework. Section 3 presents main findings. Section 4 concludes with discussions on policy implications and future research directions.

## 2   The FUND 3.9n model

We modify the FUND model to create a climate risk model tailored for individual countries. The updated national FUND model (FUND 3.9n) allows for a more granular representation of national climate risks, enabling a country-specific analysis of climate impacts and better capturing the diversity of national circumstances.[1] The revised model divides the world into 198 countries, as opposed to the original 16 regions, thereby providing a more detailed analysis of the impacts caused by climate change at the national level. The countries were selected based on those registered with the United Nations (UN) for which historical data, such as population, GDP, energy consumption, and greenhouse gas (GHG) emissions, can be established.[2] Input data were primarily sourced from international organizations, such as the United Nations (UN), the World Bank, the Organisation for Economic Co-operation and Development (OECD), the International Energy Agency (IEA), and the Food and Agriculture Organization (FAO). Climate data were drawn from the National Oceanic and Atmospheric Administration (NOAA), the Advanced Global Atmospheric Gases Experiment (AGAGE), and the Intergovernmental Panel on Climate Change (IPCC).[3] These inputs serve as initial values for calibration and projections.

The parameters of the impact module were recalibrated as follows. Unlike most other integrated assessment models, in FUND, vulnerability to climate change varies with development (Schelling 1984). Assuming ergodicity, this temporal pattern is imposed spatially. For instance, the ability to prevent infectious diseases is assumed to vary with per capita income. The assumed income elasticity is used to impute the impact in all countries that together from one of the 16 regions in FUND3.9. The national parameters for FUND3.9n are then rescaled so that the sum of the national impacts add up to the regional impact in the original model for 2.5°C global warming. Many of the impacts are part driven by an income elasticity, but we also take exposure into account. Some health impacts are specific to age groups or to cities. The impact of sea level rise depends on the length of coast and the extent of wetlands. These factors are taken into account too. The imputation algorithm is therefore different for each impact, but the principles are the same: Ergodicity plus rescaling to match the benchmark. Because the model is non-linear, national and regional impacts differ for any warming except 2.5°C.

---

[1] The model code is available at github ().

[2] The total figures for individual countries listed account for the majority of the global totals across key indicators. For instance, the combined $CO_2$ emissions from energy consumption for the 198 countries listed constitute over 99% of the global total emissions.

[3] In specific, the historical population, GDP, $CO_2$, $CH_4$, and $N_2O$ emissions, as well as geographical area by country were obtained from the World Development Indicators (https://databank.worldbank.org/source/world-development-indicators). Data on primary energy consumption were sourced from the OECD and the IEA, while urbanization rates were provided by the UN. Information on wetland and dryland areas was gathered from the FAO. National mean temperatures were supplied by Berkeley Earth (http://berkeleyearth.org/). GHG concentration were drawn from Meinshausen et al. (2020), NOAA (https://gml.noaa.gov/ccgg/trends/data.html), and AGAGE (https://agage2.eas.gatech.edu/data_archive/). Data on global mean temperature and sea-level rise were sourced from NOAA.

The model's framework includes two components: a calibration process using data from 1750 to 2018, and a projection process forecasting trends from 2019 to 2200.[4] It consists of an economic module, a carbon emission module, a climate module, and an impact module. The economic module relies on country-specific data from 1960 to 2018 to adjust growth parameters and projects trends to 2200, following neoclassical economic growth theory. It accounts for emissions of GHGs including carbon dioxide ($CO_2$), methane ($CH_4$), nitrous oxide ($N_2O$), sulfur hexafluoride ($SF_6$), CFC11, and CFC12, as well as pollutants affecting radiative forcing such as sulfur dioxide ($SO_2$) and ozone ($O_3$). The climate module includes submodules for carbon cycles, radiative forcing, temperature responses, and sea-level rise. The model evaluates physical impacts across sectors including agriculture, cooling, heating, hurricanes, extra tropical storms, forests, water resources, cardiovascular diseases, respiratory diseases, diarrheal diseases, vector-borne diseases (i.e., dengue fever, malaria, schistosomiasis), biodiversity, and sea-level rise (i.e., dryland loss, wetland loss, protection cost, migration costs). The market impacts influence economic output, and hence economic growth, while non-market impacts affect social welfare. This multi-sector approach allows for a more detailed understanding of climate impacts, supporting more informed decision-making. For comparison, we also include a number of popular aggregate impact functions fitted to the meta-analysis of Tol (2019), as well as their Bayesian Model Average. The model is designed to produce detailed national-scale analyses, making it particularly valuable for policymakers who need to evaluate the climate risks and mitigation strategies relevant to their specific country.

The future scenario for urbanization and aging follows the SSP 3.0 scenario (SSP2), while the rates of change in emission intensities are based on the SSP_IAM_V2 scenario.[5] The savings rate is derived from the current policies scenario of NGFS (2021). The concentration for CFC11 and CFC12 utilizes the SSP2-45 scenario from Meinshausen et al. (2020). All other inputs are sourced from the original FUND model. The no-policy scenario with a zero-carbon tax calculates the SCC by allowing countries to emit GHGs without accounting for externalities.[6] Improvements in energy efficiency and carbon intensity due to technological advancement result in a decline in energy and GHG intensity over time.

---

[4] In the process of applying probabilistic population scenarios for individual countries, the model year was reduced to 2200 to enhance model stability and reduce computation time. The original FUND model produces results up to the year 2300. While shortening the model year to 2200 slightly lowers the SCC values, the main implications of this paper remain largely unchanged. In addition, when incorporating stochastic population projections, some countries meet boundary conditions for population (1,000 people) and economic size (100 US$/person) during the simulation period, resulting in extremely high or low SCC values. To mitigate the impact of these outliers, the boundary for the national SCC was set at $|SCC_n|<200$ US$/tCO$_2$.

[5] Available at https://data.ece.iiasa.ac.at/ssp. The values beyond 2100 are extrapolated following the trends up to 2100.

[6] In the FUND3.9n model, the non-cooperative SCC can be derived by applying differentiated carbon taxes across countries, while the cooperative SCC can be obtained by applying a uniform carbon tax across countries. A comparison of these approaches is left for future research.

The model incorporates uncertainty across a broad range of parameters with 1,000 Monte Carlo simulations, covering climate parameters, economic parameters, and national-specific impacts parameters. The uncertainty in future national population follows the RFF scenarios (Rennert et al., 2022).

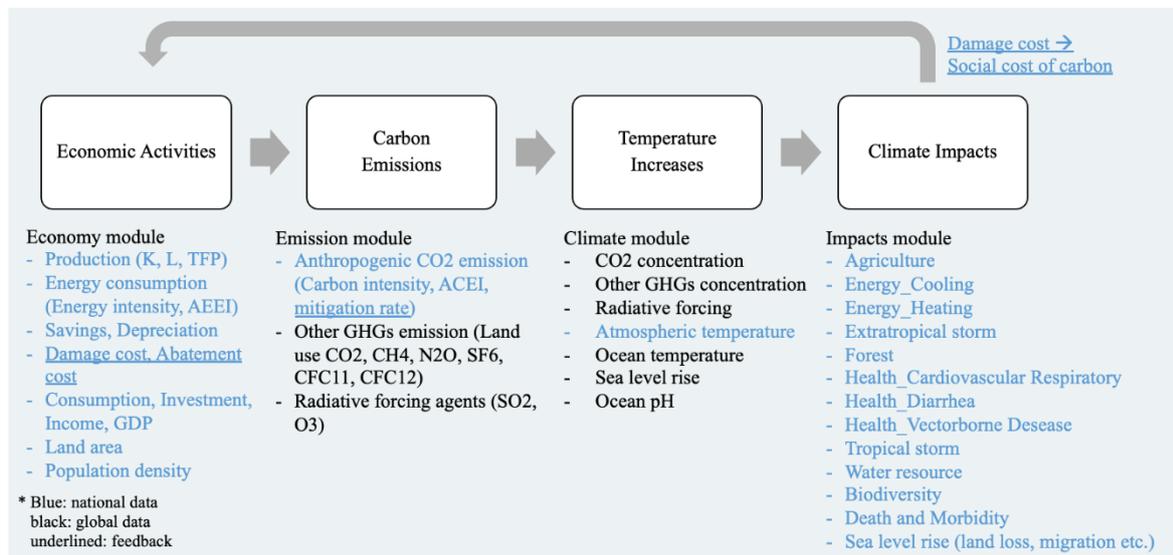

**Figure 1 FUND 3.9n model**

## 3 Results

### 3.1 National SCC

In the FUND3.9n model, sector-specific damage costs are generally formulated as increasing functions of a country's GDP and population. As a result, the NSCC generally rises with higher GDP and larger population. GDP shows a higher correlation with the NSCC (see Figure 2). It has little correlation with per capita GDP or absolute average temperature.

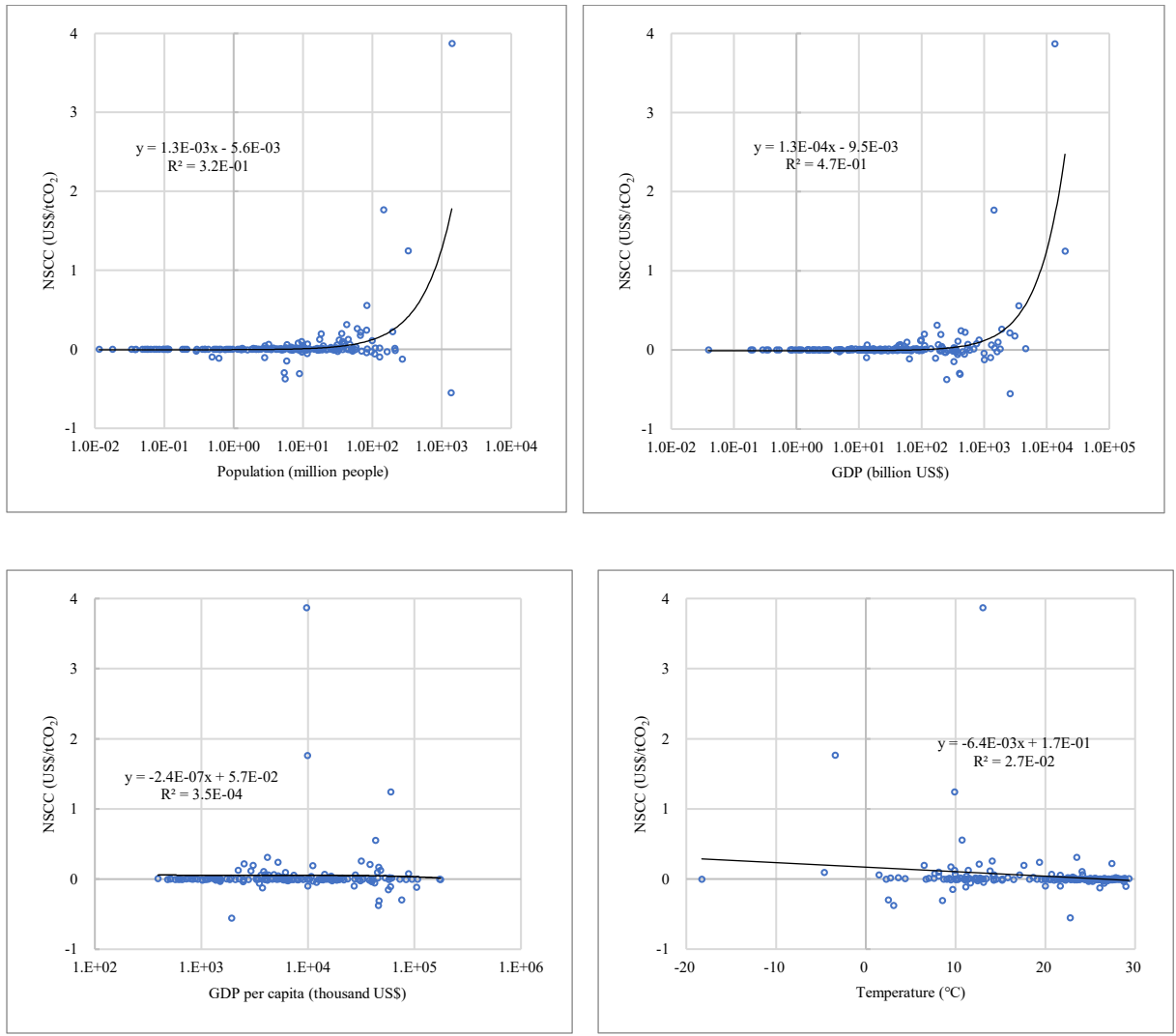

**Figure 2 NSCC according various indicators (PRTP=0.03, RRA=1)** (Top Left) Population. (Top Right) GDP. (Bottom Left) GDP per capita. (Bottom Right) Temperature.

The correlation between GDP and the NSCC varies depending on the damage function used to quantify climate impacts. The low linearity is observed when applying the damage functions proposed by Weitzman (2012), while other damage functions generally tend to exhibit stronger linear relationships.

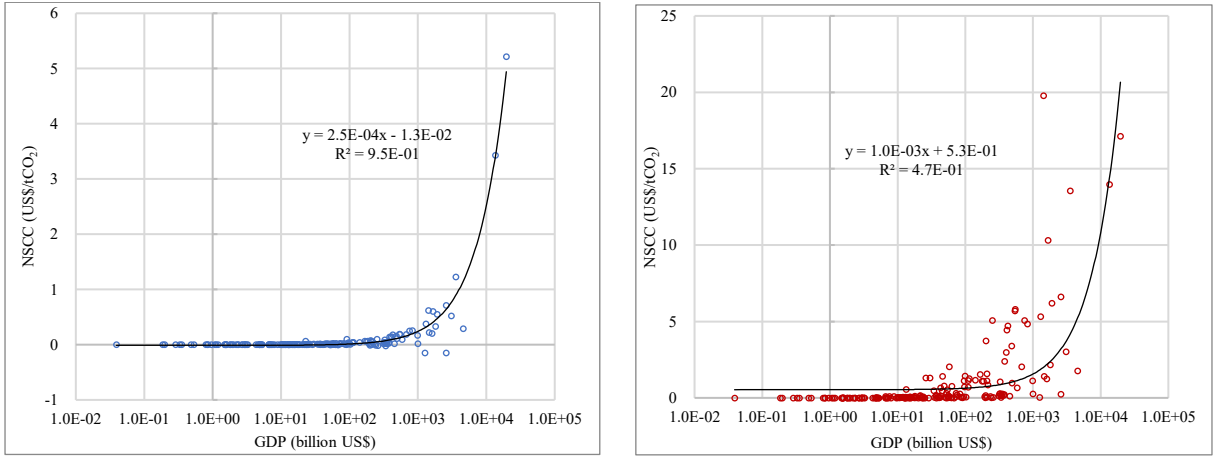

**Figure 3 NSCC according to damage function (PRTP=0.03, RRA=1).** (Left) Application of Tol (2019) damage function. (Right) Application of Weitzman (2012) damage function)

Figure 4 presents the NSCC estimates, showing that incorporating uncertainty typically leads to higher NSCC values. More detailed results for individual countries are available in the online supplementary materials.

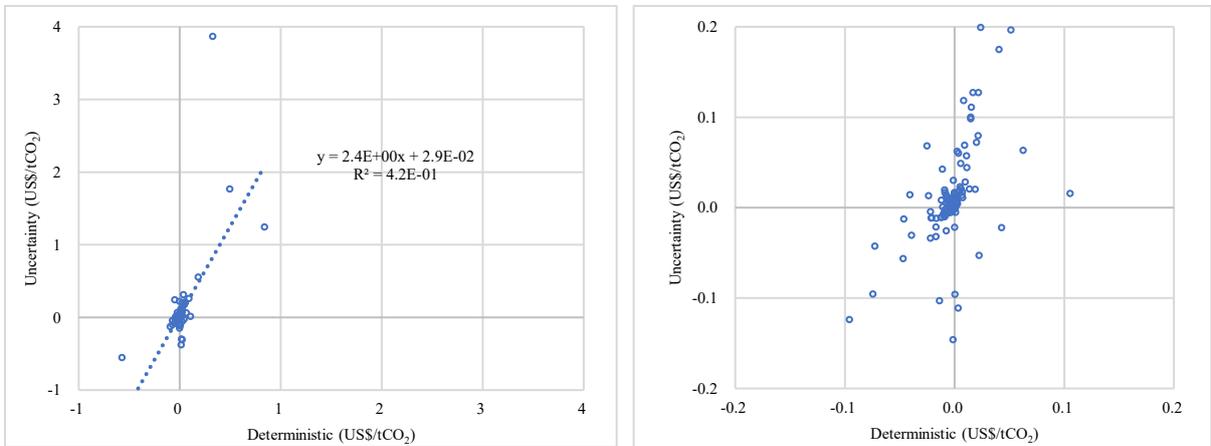

**Figure 4 NSCC: Uncertainty vs. Deterministic (PRTP=0.03, RRA=1)** (Left) Full scale (Right) Rescaled scale

Figure 5 below presents the NSCCs for the top 20 countries by GDP in the model's base year (2018).[7] The NSCCs vary significantly depending on preference parameters. Specifically, lower PRTP and RRA values are associated with higher SCCs. For instance, when PRTP is 0.03 and RRA is 1, the countries with the highest NSCCs among the top 20 countries by GDP are China (3.9 US$/tCO$_2$), Russia (1.8 US$/tCO$_2$), and the United States (1.2 US$/tCO$_2$). If PRTP decreases to 0.01, other things being equal (RRA=1), the NSCC rises significantly, with China (9.7 US$/tCO$_2$), Russia (3.8 US$/tCO$_2$), and the United States (3.8 US$/tCO$_2$). In contrast, if RRA increases to 2 (PRTP = 0.03), the NSCC values are substantially lower: China (2.9 US$/tCO$_2$), Russia (0.6 US$/tCO$_2$), and the United States (0.8 US$/tCO$_2$). India shows a negative NSCC value, as the benefits from the CO$_2$ fertilization effect outweigh the other climate impacts. For Russia, the main concern is the drying up of its water resources.

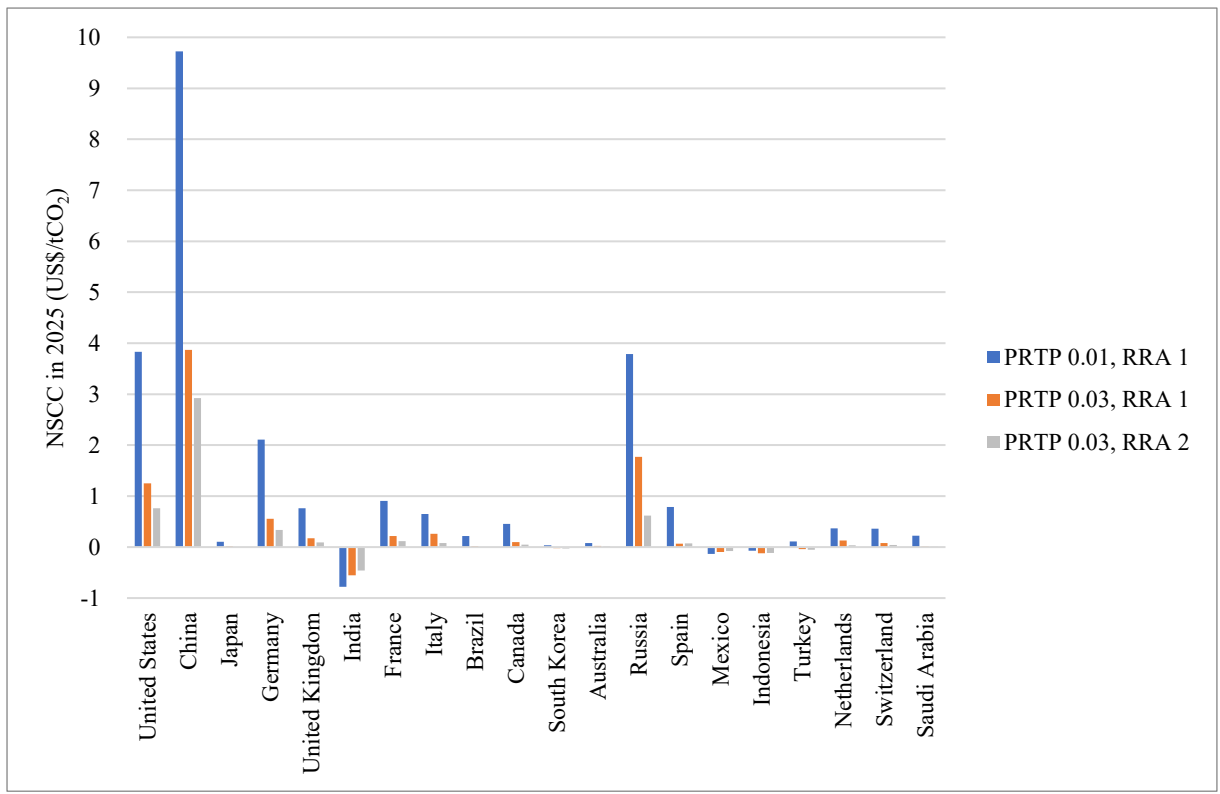

**Figure 5 NSCC according to social preferences**

---

[7] The results for all 198 countries are available in the online supplementary materials.

Figure 6 presents the distribution of the NSCC over time. The NSCC is relatively high at first in China, Russia, the United States, and several European countries. On a per capita basis, markedly positive NSCC values are observed in Russia and other transition economies, as well as Mongolia, indicating their heightened vulnerability to climate impacts. By contrast, Northern Europe and Greenland display negative values, that is, net benefits from rising temperatures, such as reduced heating costs and an expansion of agricultural opportunities. By 2100, NSCC values generally rise, although the extent of the increase differs across countries. Such variation arises from the country- and period-specific heterogeneity in sectoral impacts of climate change. For instance, climate change impacts on economic sectors including agriculture and energy consumption, tend to yield relatively large benefits until the mid-21st century, whereas damage costs increase after the latter half of the century. Given that the relative importance of each sector varies across countries and over time, so does the NSCC. For instance, South America (e.g., Brazil), the majority of Africa, Australia, and Japan see a pronounced increase in NSCC between 2025 and 2100, because the adverse impacts of climate change are projected to intensify substantially in these regions while their economies grow strongly. Conversely, some Eurasian countries including Russia, China, and India demonstrate modest change, because climate-related damages are counterbalanced by changes in the economic structure and improved adaptation.

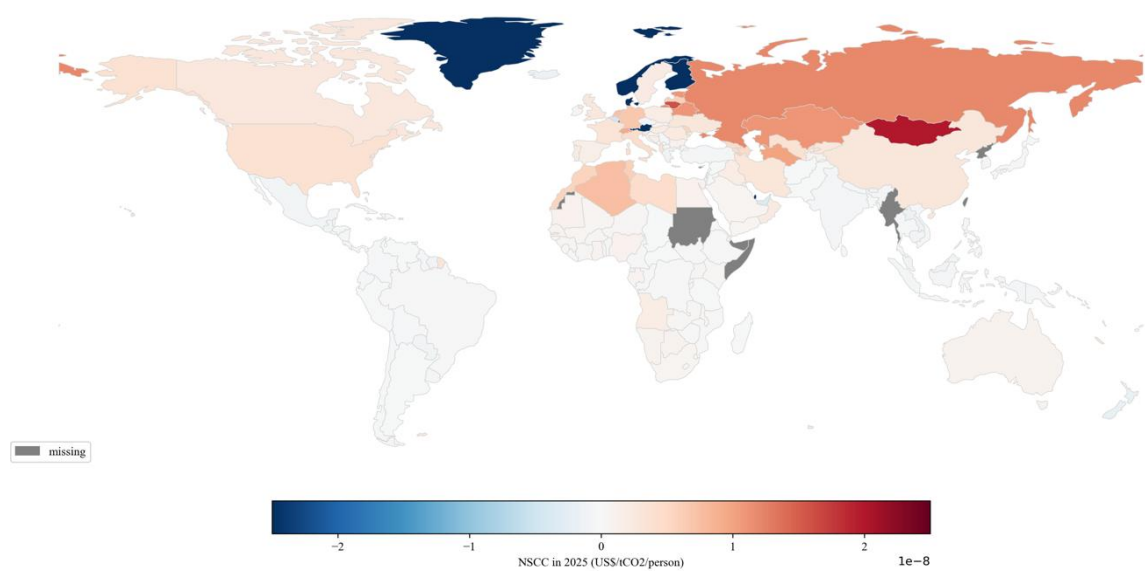

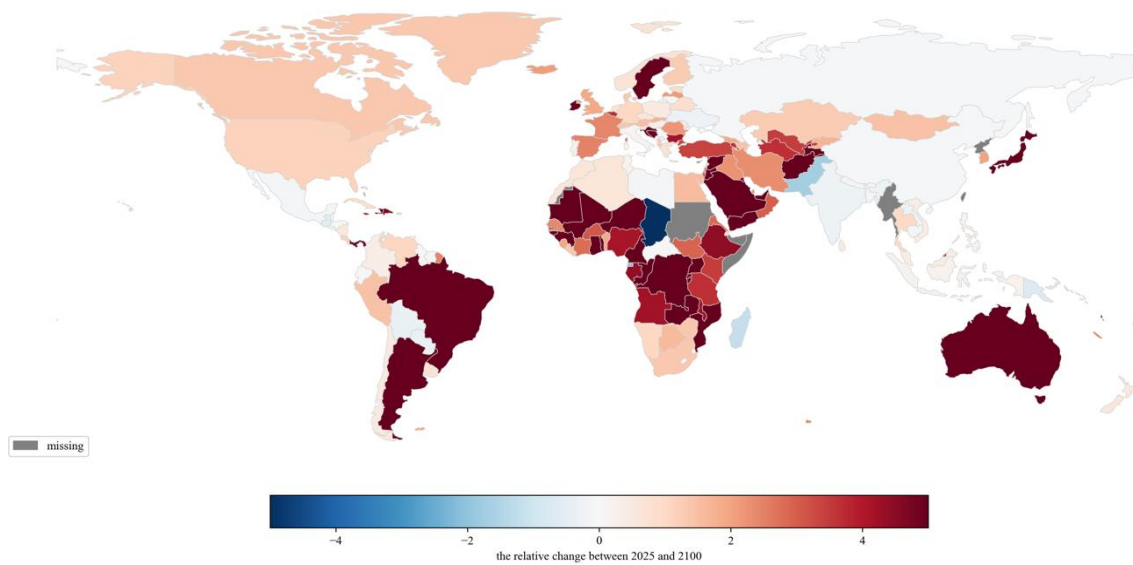

**Figure 6 NSCC distribution (PRTP=0.03, RRA=1)** (Top) The NSCC in 2025 is estimated to be $1.0 \times 10^{-8}$ US\$/tCO$_2$/person. (Bottom) the relative change with an absolute denominator in the NSCC between 2025 and 2100

Figure 7 shows the correlation between NSCC and GDP depending on income elasticity. As noted by Tol (2019), as income elasticity becomes more negative, the contribution of low-income countries to the global sum of NSCCs increases while that of high-income countries decreases. As a result, the overall linearity of the relationship is significantly reduced. The global sum of NSCCs increases from 8.4 US\$/tCO$_2$ when income elasticity is 0, to 8.9 US\$/tCO$_2$. This is because, in our parameterizations, when income elasticity is negative, the increase in the NSCC for low-income countries was generally greater than the decrease in the NSCC for high-income countries. For instance, when the income elasticity is -0.36, the NSCC of countries with a per capita GDP higher than the global average (in 2018) decreased by 0.6 US\$/tCO$_2$, while the NSCC of countries with a per capita GDP lower than the global average increased by 1.1 US\$/tCO$_2$, compared to when the income elasticity is 0.

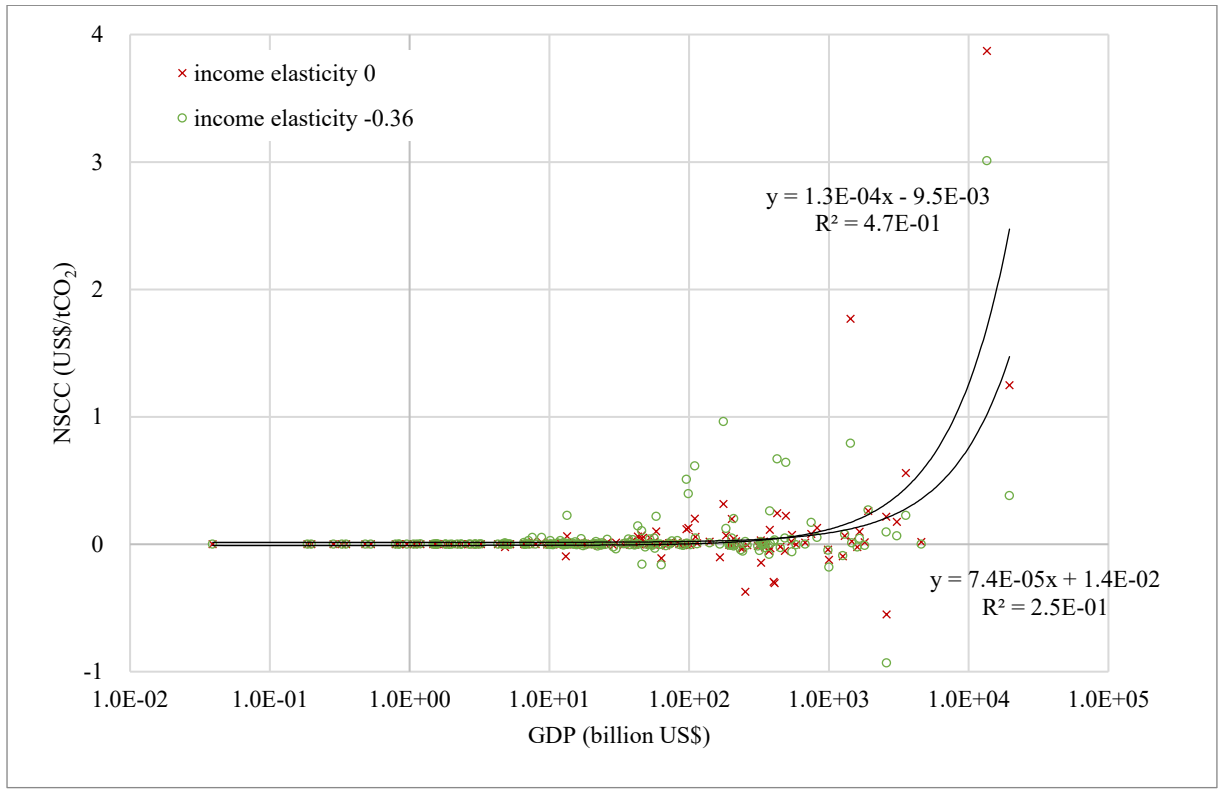

**Figure 7 NSCC according to income elasticity (PRTP=0.03, RRA=1)**

3.2 Global sum of NSCCs

The total NSCC estimated using the model is shown in Figure 8. Here, the total NSCC refers to the sum of NSCCs for all 198 countries. This is the SCC as a Lindahl price (Kelleher, 2025), viewing greenhouse gas emission reduction as a global public good. The total NSCC varies significantly depending on the impact function (see Table 1) applied. Furthermore, the total NSCC is highly sensitive to preference parameters such as the pure rate of time preference (PRTP) and relative risk aversion (RRA). In the figure, "FUND3.9n" represents the results obtained by estimating climate impacts on a country-sector basis. The other results assume a single-sector framework. When PRTP is 0.03 and RRA is 1, the reference case of this paper, the total NSCC calculated using sector-specific damage functions is 8.4 US\$/tCO$_2$. Under the single-sector assumption, the total NSCC ranges from 1.1 US\$/tCO$_2$ to 189.2 US\$/tCO$_2$ depending on the damage function. The result obtained using the damage function from Tol (2019), who applies a single-sector piecewise linear model, is 18.0 US\$/tCO$_2$. If the PRTP decreases to 0.01, other things being equal, the total NSCC estimated using sector-specific damage functions increases to 33.5 US\$/tCO$_2$, whereas it decreases to 2.7 US\$/tCO$_2$ when RRA increases to 2, other things being equal. Similar patterns are observed under the single-sector cases.

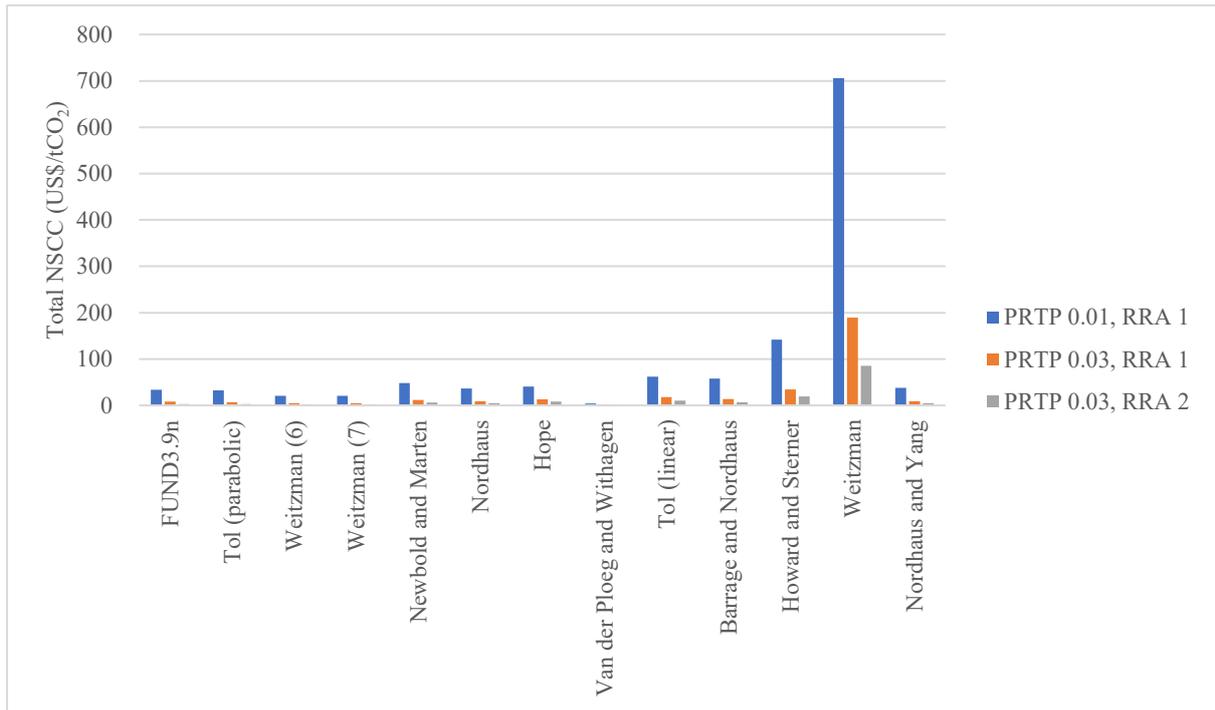

**Figure 8 Global sum of NSCCs according to impact function**

**Table 1 Impact function**

| Impact function | Functional form* | Global sum of NSCCs (US$/tCO$_2$) ** | Reference |
|---|---|---|---|
| Tol (parabolic) | $\alpha_1 T + \alpha_2 T^2$ | 6.8 | Tol (2019) |
| Weitzman (6) | $\alpha_1 T^2 + \alpha_2 T^6$ | 4.5 | Tol (2019) |
| Weitzman (7) | $\alpha_1 T^2 + \alpha_2 T^7$ | 4.7 | Tol (2019) |
| Newbold and Marten | $0\ (if\ T < \beta)$, $\alpha_1(T - \beta)\ (if\ T \geq \beta)$ | 11.5 | Tol (2019) |
| Nordhaus | $\alpha_1 T^2$ | 9.0 | Tol (2019) |
| Hope | $\alpha_1 T$ | 13.1 | Tol (2019) |
| Van der Ploeg and Withagen | $\alpha_1(\exp(T) - 1)$ | 1.1 | Tol (2019) |
| Tol (linear) | $\alpha_1 T\ (if\ T < \beta)$, $\alpha_2 + \alpha_3 T\ (if\ T \geq \beta)$ | 18.0 | Tol (2019) |
| Barrage and Nordhaus | $\alpha_1 T^2$ | 14.1 | Barrage and Nordhaus (2024) |
| Howard and Sterner | $\alpha_1 T^2$ | 34.7 | Howard and Sterner (2017) |
| Weitzman | $\alpha_1 T^2 + \alpha_2 T^{6.754}$ | 189.2 | Weitzman (2012) |
| Nordhaus and Yang | $\alpha_1 T^2$ | 9.2 | Nordhaus and Yang (1996) |
| BMA*** |  | 5.1 | This study |
| FUND3.9n | sectoral impact functions | 8.4 | This study |

Note: * $T$ is the temperature increase, $\alpha$ and $\beta$ are the parameters. For the parameters see the references.
** The results are for PRTP=0.03, and RRA=1.
*** The BMA results correspond to the 8 independent functional forms including Tol (parabolic), Weitzman (6), Weitzman (7), Newbold and Marten, Nordhaus, Hope, Van der Ploeg and Withagen, and Tol (linear). For the 8 impact functions, the parameters were calibrated based on historical climate impact data.

Table 2 illustrates that sectoral disaggregation and the inclusion of uncertainty significantly influence the estimated total NSCC. Across all preference parameter settings, sectoral disaggregation leads to lower NSCC and incorporating uncertainty leads to higher NSCC estimates in both the sectoral (FUND3.9n) and the aggregated (Tol, 2019) models.

**Table 2 Global sum of NSCCs (US$/tCO$_2$)**

|  | Preferences | | Impact function | |
|---|---|---|---|---|
|  | PRTP | RRA | FUND3.9n | Tol (2019) |
| Uncertainty | 0.01 | 1 | 33.5 | 61.9 |
|  | 0.03 | 1 | 8.4 | 18.0 |
|  | 0.03 | 2 | 2.7 | 10.6 |
| Deterministic | 0.01 | 1 | 8.4 | 51.3 |
|  | 0.03 | 1 | 1.1 | 14.4 |
|  | 0.03 | 2 | -0.0 | 8.3 |

## 4 Conclusions

Addressing global environmental challenges necessitates strategic interventions, particularly through the implementation of carbon pricing. The SCC is a crucial metric for evaluating the economic impacts of climate change. Previous models have primarily focused on global or regional scales, often failing to capture the nuances of individual countries' economic and social contexts. Furthermore, in practice, it is difficult to impose a harmonized carbon price based on the global SCC. This has been clearly demonstrated by the climate change negotiations over the past few decades.

This paper builds on the existing FUND model to present a modified version that estimates NSCC for 198 countries. This approach allows for a more granular assessment of the costs of climate change by incorporating sector-specific impacts, heterogeneity in country-level vulnerabilities, and uncertainty. This is our main contribution. The results highlight the importance of incorporating sectoral vulnerability and uncertainty in climate policy, particularly for carbon pricing and adaptation strategies. The NSCC varies substantially when sector-specific climate impacts and uncertainty are incorporated, as opposed to modeling each country as a single aggregate sector. These differences arise from country-level heterogeneity in the relative importance of sectors, their sensitivity to climate change, and the degree to which uncertainty amplifies sectoral damages. Under the parameterization used in this paper, accounting for uncertainty leads to a considerably larger increase in the global sum of NSCCs than accounting for sectoral differentiation alone. Compared to previous studies, the contribution of high-income countries to the global sum of NSCC tends to decrease, while the contribution from low-income countries increases.

The NSCC estimates presented here can inform climate policy design, ensuring that national policies are more aligned with the unique climate risks and economic structures of individual countries. For example, they can be used to design more targeted mitigation measures, allocate resources for national adaptation planning, and contribute to international burden-sharing negotiations. Additionally, NSCCs can help ensure that climate policies reflect local economic and environmental conditions, promoting fairer and more effective global climate governance.

While this paper does not advocate for country-specific carbon prices (Liu and Raftery, 2021), NSCC estimates can still serve as valuable indicators in domestic climate decision-making. For instance, they can guide national adaptation planning and budgeting by highlighting sectoral impacts at the country level. They may also serve as a lower bound in determining the ambition of national mitigation efforts, with the global SCC acting as an upper bound. Moreover, NSCCs could be useful in international negotiations over burden-sharing and emissions allocation. Since they reflect the marginal damage of one additional tonne of carbon emitted globally but experienced locally, they offer a damage-based perspective that can complement traditional allocation principles such as the polluter pays principle, per capita emissions, or income-based fairness (Höhne et al., 2014; Ohndorf et al., 2015; Pattanayak and Kumar, 2015; Kornek et al., 2017; Sheriff, 2019; Fyson et al., 2020; Kesternich et al., 2021). Exploring the integration of impact-based metrics into future burden-sharing frameworks remains an area for further research.

Future research should focus on refining the model's sectoral parameters for specific countries and integrating more detailed climate impact scenarios. Further developments could also include incorporating the urban heat island effect (Estrada et al., 2017) and other localized impacts to better capture the realities of climate change at the urban level. This would allow for a deeper understanding of localized climate risks and provide more actionable insights for national and international climate policy. In addition, national preferences can be incorporated into the model, providing the ability to account for country-specific priorities and socio-economic conditions in the evaluation of climate risks and mitigation strategies (Dong et al., 2024). There has been growing demand from the financial and industrial sectors for models that comprehensively assess how climate risks propagate throughout the entire economic system. In the long term, it may be worth considering incorporating the strengths of the national FUND model into the development of an integrated climate-economy-finance model to address this need. Alternatively, it may be possible to link it with other economic models. For instance, an approach would involve developing an algorithm that effectively integrates the strengths of a computable general equilibrium (CGE) model, which reflects the industrial and economic structure of individual countries, with the strengths of the FUND model, which captures the interactions between climate and the economy (e.g., Hsiang et al., 2017; Ciscar et al., 2019; Neumann et al., 2020).

## Acknowledgements

This work was supported by Korea Environment Industry & Technology Institute (KEITI) through Climate Change R&D Project for New Climate Regime, funded by Korea Ministry of Environment (MOE)(RS-2023-00218794). We are also grateful to Jong-rak Baek and Jeewon Son for their support in data collection.

## Online Supplementary Information

The NSCCs for 198 countries are shown in Tables A and B.

**Table A. NSCCs for 198 countries in 2025 (unit: US$/tCO$_2$)**

| Country | PRTP = 0.01, RRA = 1 | | PRTP = 0.03, RRA = 1 | | PRTP = 0.03, RRA = 2 | |
|---|---|---|---|---|---|---|
| | Uncertainty | Deterministic | Uncertainty | Deterministic | Uncertainty | Deterministic |
| Afghanistan | 0.033 | -0.009 | -0.001 | -0.008 | 0.000 | -0.007 |
| Albania | 0.087 | -0.003 | 0.011 | -0.003 | 0.869 | -0.003 |
| Algeria | 1.113 | 0.115 | 0.315 | 0.036 | 0.461 | 0.021 |
| Andorra | 0.000 | 0.000 | 0.000 | 0.000 | 0.000 | 0.000 |
| Angola | 0.147 | 0.035 | 0.057 | 0.011 | 0.022 | 0.006 |
| Antigua and Barbuda | 0.000 | 0.000 | 0.000 | 0.000 | 0.000 | 0.000 |
| Argentina | 0.031 | 0.025 | -0.002 | -0.002 | -0.006 | -0.006 |
| Armenia | 0.073 | 0.004 | 0.013 | 0.000 | -0.657 | -0.001 |
| Aruba | 0.000 | 0.000 | 0.000 | 0.000 | 0.000 | 0.000 |
| Australia | 0.081 | 0.004 | 0.023 | 0.005 | 0.014 | 0.004 |
| Austria | 0.243 | 0.066 | -0.305 | 0.023 | 0.036 | 0.014 |
| Azerbaijan | 0.442 | 0.028 | 0.049 | 0.006 | 0.165 | 0.002 |
| Bahamas | 0.002 | 0.001 | 0.001 | 0.000 | 0.000 | 0.000 |
| Bahrain | -0.005 | 0.002 | -0.005 | 0.001 | 0.002 | 0.000 |
| Bangladesh | -0.027 | -0.059 | -0.030 | -0.039 | -0.060 | -0.033 |
| Barbados | -0.087 | 0.000 | -0.022 | 0.000 | 0.251 | 0.000 |

| Country | | | | | | |
|---|---|---|---|---|---|---|
| Belarus | 0.286 | 0.052 | 0.100 | 0.015 | 1.086 | 0.008 |
| Belgium | 0.304 | 0.067 | -0.053 | 0.022 | 0.041 | 0.013 |
| Belize | 0.000 | 0.000 | 0.000 | 0.000 | 0.000 | 0.000 |
| Benin | 0.030 | -0.002 | 0.003 | -0.003 | 0.078 | -0.003 |
| Bermuda | 0.000 | 0.000 | 0.000 | 0.000 | 0.000 | 0.000 |
| Bhutan | -0.001 | -0.001 | -0.001 | -0.001 | -0.001 | -0.001 |
| Bolivia | -0.003 | -0.006 | -0.003 | -0.004 | -0.003 | -0.003 |
| Bosnia and Herzegovina | 0.031 | 0.002 | 0.004 | 0.000 | 0.408 | 0.000 |
| Botswana | 0.008 | 0.002 | 0.002 | 0.001 | 0.001 | 0.000 |
| Brazil | 0.216 | 0.000 | 0.015 | -0.041 | -0.012 | -0.042 |
| Brunei | 0.002 | 0.001 | 0.000 | 0.000 | 0.000 | 0.000 |
| Bulgaria | -0.029 | 0.005 | 0.007 | 0.000 | 0.075 | -0.001 |
| Burkina Faso | 0.039 | 0.000 | 0.006 | -0.002 | 0.012 | -0.002 |
| Burundi | 0.011 | -0.002 | 0.000 | -0.002 | 0.001 | -0.002 |
| Coted'Ivoire | 0.042 | -0.006 | 0.005 | -0.007 | -0.088 | -0.007 |
| Cambodia | -0.007 | -0.013 | -0.010 | -0.009 | -0.008 | -0.008 |
| Cameroon | 0.048 | -0.003 | 0.004 | -0.005 | 0.003 | -0.005 |
| Canada | 0.456 | 0.089 | 0.098 | 0.014 | 0.050 | 0.003 |
| Cape Verde | 0.001 | 0.000 | 0.000 | 0.000 | 0.097 | 0.000 |
| Central African Republic | 0.002 | -0.001 | 0.000 | -0.001 | 0.000 | -0.001 |
| Chad | 0.013 | -0.003 | 0.000 | -0.003 | 0.000 | -0.003 |
| Chile | -0.004 | -0.015 | -0.007 | -0.010 | -0.006 | -0.008 |
| China | 9.723 | 2.772 | 3.871 | 0.324 | 2.925 | 0.019 |
| Colombia | -0.004 | -0.020 | -0.012 | -0.017 | -0.012 | -0.015 |
| Comoros | 0.004 | 0.000 | 0.000 | 0.000 | 0.000 | 0.000 |
| Congo [DRC] | 0.029 | 0.001 | 0.005 | -0.003 | 0.002 | -0.003 |
| Congo [Republic] | 0.016 | -0.007 | -0.002 | -0.005 | -0.012 | -0.004 |
| Costa Rica | 0.002 | -0.007 | -0.003 | -0.005 | -0.003 | -0.004 |
| Croatia | -0.149 | 0.009 | 0.003 | 0.001 | 0.165 | 0.000 |
| Cuba | 0.006 | -0.014 | -0.006 | -0.010 | -0.006 | -0.008 |
| Cyprus | 0.005 | 0.000 | 0.001 | 0.000 | 0.000 | -0.001 |
| Czech Republic | -0.195 | 0.014 | -0.005 | -0.002 | 0.105 | -0.004 |
| Denmark | 0.086 | 0.013 | -0.146 | -0.002 | 0.006 | -0.004 |
| Djibouti | 0.006 | 0.001 | 0.001 | 0.000 | -0.005 | 0.000 |
| Dominica | 0.000 | 0.000 | 0.000 | 0.000 | 0.000 | 0.000 |
| Dominican Republic | 0.007 | -0.005 | -0.001 | -0.004 | -0.002 | -0.003 |

| Country | | | | | | |
|---|---|---|---|---|---|---|
| Ecuador | -0.002 | -0.003 | -0.005 | -0.004 | -0.005 | -0.004 |
| Egypt | 0.528 | 0.085 | 0.111 | 0.015 | 0.045 | 0.002 |
| El Salvador | 0.000 | -0.003 | -0.001 | -0.002 | -0.001 | -0.002 |
| Equatorial Guinea | 0.004 | -0.003 | -0.002 | -0.002 | -0.002 | -0.002 |
| Eritrea | 0.002 | 0.000 | 0.000 | 0.000 | 0.000 | 0.000 |
| Estonia | -0.245 | 0.020 | 0.014 | 0.007 | -0.413 | 0.004 |
| Ethiopia | 0.118 | 0.002 | 0.017 | -0.009 | 0.017 | -0.010 |
| Faroe Islands | 0.000 | 0.000 | 0.000 | 0.000 | 0.000 | 0.000 |
| Fiji | 0.004 | -0.001 | 0.000 | -0.001 | -0.005 | -0.001 |
| Finland | 0.355 | 0.044 | -0.373 | 0.013 | -0.068 | 0.007 |
| France | 0.907 | 0.224 | 0.215 | 0.052 | 0.120 | 0.023 |
| Gabon | 0.006 | 0.001 | 0.001 | 0.000 | 0.001 | 0.000 |
| Gambia | 0.006 | 0.000 | 0.001 | 0.000 | 0.010 | 0.000 |
| Georgia | 0.118 | 0.005 | 0.016 | -0.001 | -0.041 | -0.001 |
| Germany | 2.108 | 0.539 | 0.557 | 0.182 | 0.338 | 0.112 |
| Ghana | 0.142 | -0.001 | 0.020 | -0.009 | -0.068 | -0.010 |
| Greece | 0.025 | -0.028 | -0.011 | -0.021 | -0.013 | -0.019 |
| Greenland | 0.000 | 0.000 | 0.000 | 0.000 | 0.000 | 0.000 |
| Grenada | 0.000 | 0.000 | 0.000 | 0.000 | 0.000 | 0.000 |
| Guatemala | -0.011 | -0.020 | -0.011 | -0.013 | -0.009 | -0.010 |
| Guinea | 0.016 | 0.001 | 0.003 | -0.001 | 0.003 | -0.001 |
| Guinea-Bissau | 0.005 | 0.000 | 0.000 | 0.000 | 0.001 | 0.000 |
| Guyana | 0.000 | 0.000 | 0.000 | 0.000 | 0.000 | 0.000 |
| Haiti | 0.007 | -0.005 | -0.001 | -0.004 | -0.006 | -0.003 |
| Honduras | -0.001 | -0.005 | -0.002 | -0.003 | -0.002 | -0.003 |
| Hong Kong | 0.060 | 0.019 | 0.018 | 0.007 | 0.013 | 0.005 |
| Hungary | 0.122 | 0.023 | 0.019 | 0.004 | 0.015 | 0.001 |
| Iceland | 0.001 | -0.001 | 0.000 | -0.001 | -0.001 | -0.001 |
| India | -0.776 | -0.901 | -0.551 | -0.574 | -0.459 | -0.470 |
| Indonesia | -0.068 | -0.105 | -0.124 | -0.096 | -0.117 | -0.087 |
| Iran | -0.035 | -0.041 | 0.244 | -0.052 | 0.005 | -0.047 |
| Iraq | 0.398 | -0.027 | 0.068 | -0.025 | 0.026 | -0.022 |
| Ireland | 0.041 | -0.008 | 0.001 | -0.011 | -0.004 | -0.010 |
| Isle of Man | 0.000 | 0.000 | 0.000 | 0.000 | 0.000 | 0.000 |
| Israel | 0.011 | -0.034 | -0.011 | -0.022 | -0.014 | -0.017 |
| Italy | 0.649 | 0.278 | 0.261 | 0.088 | 0.077 | 0.047 |

| Country | | | | | | |
|---|---|---|---|---|---|---|
| Jamaica | 0.001 | -0.001 | 0.000 | -0.001 | 0.000 | -0.001 |
| Japan | 0.107 | 0.177 | 0.016 | 0.105 | 0.004 | 0.078 |
| Jordan | 0.001 | -0.003 | -0.001 | -0.002 | -0.001 | -0.002 |
| Kazakhstan | -0.384 | 0.167 | 0.197 | 0.051 | 0.192 | 0.031 |
| Kenya | 0.133 | -0.002 | 0.015 | -0.008 | -0.002 | -0.008 |
| Kiribati | 0.000 | 0.000 | 0.000 | 0.000 | 0.000 | 0.000 |
| Kosovo | 0.000 | 0.000 | 0.000 | 0.000 | 0.000 | 0.000 |
| Kuwait | 0.026 | 0.006 | 0.007 | 0.002 | 0.005 | 0.001 |
| Kyrgyzstan | 0.106 | 0.003 | 0.017 | 0.000 | 0.019 | -0.001 |
| Laos | -0.009 | -0.013 | -0.009 | -0.009 | -0.007 | -0.008 |
| Latvia | -0.060 | 0.022 | 0.011 | 0.007 | -0.139 | 0.004 |
| Lebanon | 0.002 | -0.007 | -0.002 | -0.004 | -0.003 | -0.005 |
| Lesotho | 0.001 | 0.000 | 0.000 | 0.000 | 0.000 | 0.000 |
| Liberia | 0.004 | 0.000 | 0.001 | 0.000 | 0.014 | 0.000 |
| Libya | 0.192 | 0.005 | 0.030 | -0.001 | 0.069 | -0.002 |
| Liechtenstein | -0.027 | 0.000 | -0.001 | 0.000 | 0.186 | 0.000 |
| Lithuania | -0.287 | 0.034 | 0.045 | 0.011 | -0.013 | 0.006 |
| Luxembourg | 0.070 | 0.009 | -0.111 | 0.003 | 0.100 | 0.002 |
| Macau | 0.004 | -0.005 | -0.002 | -0.004 | -0.002 | -0.003 |
| Macedonia [FYROM] | 0.004 | 0.001 | 0.001 | 0.000 | 0.000 | 0.000 |
| Madagascar | 0.011 | -0.005 | -0.002 | -0.005 | -0.001 | -0.004 |
| Malawi | 0.039 | 0.005 | 0.010 | 0.001 | 0.011 | 0.000 |
| Malaysia | -0.005 | 0.003 | -0.025 | -0.008 | -0.021 | -0.010 |
| Maldives | 0.004 | 0.000 | 0.000 | 0.000 | 0.039 | 0.000 |
| Mali | 0.086 | -0.003 | 0.008 | -0.005 | 0.017 | -0.004 |
| Malta | -0.235 | 0.001 | -0.096 | 0.000 | 0.292 | 0.000 |
| Marshall Islands | 0.000 | 0.000 | 0.000 | 0.000 | 0.000 | 0.000 |
| Mauritania | 0.024 | 0.001 | 0.004 | 0.000 | -0.005 | 0.000 |
| Mauritius | 0.003 | 0.000 | 0.001 | 0.000 | -0.001 | 0.000 |
| Mexico | -0.134 | -0.108 | -0.095 | -0.075 | -0.079 | -0.064 |
| Micronesia | 0.000 | 0.000 | 0.000 | 0.000 | 0.000 | 0.000 |
| Moldova | 0.080 | 0.006 | 0.015 | 0.001 | -0.597 | 0.000 |
| Monaco | -0.009 | 0.000 | 0.000 | 0.000 | 0.019 | 0.000 |
| Mongolia | 0.226 | 0.010 | 0.063 | 0.002 | -1.781 | 0.001 |
| Montenegro | 0.000 | 0.000 | 0.000 | 0.000 | 0.000 | 0.000 |
| Morocco | 0.937 | 0.076 | 0.200 | 0.024 | -0.924 | 0.014 |

| Country | | | | | | |
|---|---|---|---|---|---|---|
| Mozambique | 0.016 | -0.003 | 0.000 | -0.004 | 0.000 | -0.003 |
| Namibia | 0.011 | 0.001 | 0.002 | 0.000 | -0.091 | 0.000 |
| Nepal | -0.015 | -0.019 | -0.011 | -0.012 | -0.009 | -0.010 |
| Netherlands | 0.367 | 0.080 | 0.127 | 0.017 | 0.035 | 0.006 |
| New Zealand | -0.008 | -0.018 | -0.008 | -0.010 | -0.007 | -0.008 |
| Nicaragua | 0.001 | -0.005 | -0.002 | -0.003 | -0.002 | -0.002 |
| Niger | 0.051 | -0.003 | 0.005 | -0.004 | 0.009 | -0.004 |
| Nigeria | 0.652 | 0.113 | 0.223 | -0.002 | 0.236 | -0.022 |
| Norway | 0.211 | 0.053 | -0.295 | 0.017 | 0.031 | 0.011 |
| Oman | -0.148 | -0.008 | 0.010 | -0.006 | 0.016 | -0.005 |
| Pakistan | 0.202 | -0.062 | -0.012 | -0.046 | 0.022 | -0.039 |
| Palau | 0.000 | 0.000 | 0.000 | 0.000 | 0.000 | 0.000 |
| Palestinian Territories | 0.000 | 0.000 | 0.000 | 0.000 | 0.000 | 0.000 |
| Panama | 0.005 | -0.002 | 0.000 | -0.002 | 0.000 | -0.001 |
| Papua New Guinea | -0.001 | -0.003 | -0.004 | -0.003 | -0.004 | -0.003 |
| Paraguay | -0.009 | -0.008 | -0.006 | -0.006 | -0.005 | -0.005 |
| Peru | 0.004 | -0.001 | -0.003 | -0.005 | -0.004 | -0.005 |
| Philippines | -0.021 | -0.053 | -0.056 | -0.047 | -0.047 | -0.044 |
| Poland | 0.307 | 0.076 | 0.072 | 0.020 | 0.038 | 0.009 |
| Portugal | 0.203 | 0.023 | 0.021 | 0.006 | 0.033 | 0.002 |
| Puerto Rico | 0.010 | 0.001 | 0.002 | -0.001 | 0.001 | -0.001 |
| Qatar | 0.124 | -0.007 | -0.103 | -0.014 | 0.005 | -0.013 |
| Romania | -0.057 | 0.006 | 0.043 | -0.011 | 0.007 | -0.012 |
| Russia | 3.784 | 1.469 | 1.768 | 0.492 | 0.617 | 0.311 |
| Rwanda | 0.013 | -0.002 | 0.000 | -0.002 | 0.001 | -0.002 |
| São Tomé and Príncipe | 0.000 | 0.000 | 0.000 | 0.000 | 0.000 | 0.000 |
| Saint Kitts and Nevis | 0.000 | 0.000 | 0.000 | 0.000 | 0.000 | 0.000 |
| Saint Lucia | 0.001 | 0.000 | 0.000 | 0.000 | 0.000 | 0.000 |
| Saint Vincent and the Grenadines | 0.000 | 0.000 | 0.000 | 0.000 | 0.000 | 0.000 |
| Samoa | 0.000 | 0.000 | 0.000 | 0.000 | 0.000 | 0.000 |
| San Marino | 0.002 | 0.000 | 0.000 | 0.000 | -0.008 | 0.000 |
| Saudi Arabia | 0.226 | -0.022 | 0.013 | -0.024 | 0.012 | -0.022 |
| Senegal | 0.068 | 0.004 | 0.012 | 0.000 | 0.008 | 0.000 |
| Serbia | 0.000 | 0.000 | 0.000 | 0.000 | 0.000 | 0.000 |
| Seychelles | 0.000 | 0.000 | 0.000 | 0.000 | 0.000 | 0.000 |
| Sierra Leone | 0.015 | -0.001 | 0.002 | -0.001 | 0.004 | -0.001 |

| Country | | | | | | |
|---|---|---|---|---|---|---|
| Singapore | 0.064 | 0.031 | 0.021 | 0.013 | 0.014 | 0.010 |
| Slovakia | -0.048 | 0.010 | 0.005 | 0.002 | 0.010 | 0.001 |
| Slovenia | 0.025 | 0.006 | 0.006 | 0.002 | 0.004 | 0.001 |
| Solomon Islands | 0.000 | 0.000 | 0.000 | 0.000 | 0.000 | 0.000 |
| South Africa | 0.113 | 0.033 | 0.029 | 0.009 | 0.016 | 0.004 |
| South Korea | 0.036 | 0.115 | -0.022 | 0.043 | -0.026 | 0.024 |
| Spain | 0.789 | 0.193 | 0.064 | 0.062 | 0.071 | 0.035 |
| Sri Lanka | -0.006 | -0.012 | -0.007 | -0.009 | -0.009 | -0.008 |
| Sudan | 0.043 | 0.007 | 0.008 | -0.001 | 0.000 | -0.002 |
| Suriname | 0.000 | -0.001 | 0.000 | -0.001 | 0.000 | 0.000 |
| Swaziland | 0.000 | 0.000 | 0.000 | 0.000 | 0.000 | 0.000 |
| Sweden | 0.290 | 0.072 | 0.020 | 0.019 | 0.037 | 0.009 |
| Switzerland | 0.362 | 0.083 | 0.080 | 0.021 | 0.044 | 0.010 |
| Syria | 0.010 | -0.006 | -0.001 | -0.004 | -0.002 | -0.003 |
| Tajikistan | 0.059 | 0.000 | 0.008 | -0.002 | 0.009 | -0.002 |
| Tanzania | 0.100 | -0.026 | -0.004 | -0.022 | 0.000 | -0.019 |
| Thailand | 0.013 | -0.007 | -0.021 | -0.017 | -0.024 | -0.018 |
| Timor-Leste | -0.002 | -0.002 | -0.001 | -0.001 | -0.001 | -0.001 |
| Togo | 0.014 | -0.001 | 0.001 | -0.002 | 0.002 | -0.001 |
| Tonga | 0.000 | 0.000 | 0.000 | 0.000 | 0.004 | 0.000 |
| Trinidad and Tobago | 0.002 | 0.001 | 0.001 | 0.000 | 0.000 | 0.000 |
| Tunisia | 0.426 | 0.027 | 0.069 | 0.009 | -0.305 | 0.005 |
| Turkey | 0.111 | -0.084 | -0.042 | -0.073 | -0.051 | -0.065 |
| Turkmenistan | 0.104 | 0.022 | 0.060 | 0.003 | 0.805 | 0.000 |
| Tuvalu | 0.000 | 0.000 | 0.000 | 0.000 | 0.000 | 0.000 |
| Uganda | 0.104 | -0.010 | 0.008 | -0.012 | 0.018 | -0.011 |
| Ukraine | 0.650 | 0.080 | 0.127 | 0.022 | -0.298 | 0.014 |
| United Arab Emirates | 0.279 | -0.017 | -0.034 | -0.022 | 0.004 | -0.022 |
| United Kingdom | 0.761 | 0.185 | 0.175 | 0.040 | 0.090 | 0.015 |
| United States | 3.833 | 2.520 | 1.248 | 0.839 | 0.760 | 0.509 |
| Uruguay | -0.001 | -0.001 | -0.001 | -0.001 | -0.001 | -0.001 |
| Uzbekistan | 0.608 | 0.047 | 0.119 | 0.008 | -2.118 | 0.002 |
| Vanuatu | 0.000 | 0.000 | 0.000 | 0.000 | -0.001 | 0.000 |
| Venezuela | 0.004 | -0.010 | -0.006 | -0.010 | -0.006 | -0.009 |
| Vietnam | 0.007 | 0.015 | -0.032 | -0.017 | -0.030 | -0.022 |
| Yemen | 0.134 | -0.008 | 0.013 | -0.007 | -0.026 | -0.006 |

| Country | | |
|---------|---|---|
| Zambia | 0.024 | 0.003 | 0.006 | 0.000 | -0.002 | -0.001 |
| Zimbabwe | 0.014 | 0.001 | 0.002 | -0.001 | 0.001 | -0.001 |

**Table B. NSCCs for 198 countries in 2100 (unit: US$/tCO$_2$)**

| Country | PRTP = 0.01, RRA = 1 | | PRTP = 0.03, RRA = 1 | | PRTP = 0.03, RRA = 2 | |
|---------|---|---|---|---|---|---|
| | Uncertainty | Deterministic | Uncertainty | Deterministic | Uncertainty | Deterministic |
| Afghanistan | 0.147 | -0.021 | 0.033 | -0.023 | 0.070 | -0.021 |
| Albania | 0.056 | -0.002 | 0.025 | -0.002 | 0.290 | -0.002 |
| Algeria | 0.891 | 0.174 | 0.503 | 0.075 | 0.339 | 0.046 |
| Andorra | 0.000 | 0.000 | 0.000 | 0.000 | -0.013 | 0.000 |
| Angola | 0.739 | 0.177 | 0.300 | 0.071 | 0.235 | 0.040 |
| Antigua and Barbuda | 0.000 | 0.000 | 0.000 | 0.000 | 0.000 | 0.000 |
| Argentina | 0.065 | 0.057 | 0.025 | 0.026 | 0.014 | 0.015 |
| Armenia | 0.261 | 0.004 | 0.059 | 0.000 | -0.561 | 0.000 |
| Aruba | 0.000 | 0.000 | 0.000 | 0.000 | 0.000 | 0.000 |
| Australia | 0.274 | 0.101 | 0.148 | 0.070 | 0.108 | 0.057 |
| Austria | 0.417 | 0.101 | 0.158 | 0.049 | 0.132 | 0.033 |
| Azerbaijan | 0.349 | 0.025 | 0.121 | 0.008 | 0.190 | 0.004 |
| Bahamas | 0.004 | 0.002 | 0.002 | 0.001 | 0.001 | 0.000 |
| Bahrain | 0.015 | 0.007 | 0.008 | 0.004 | 0.006 | 0.003 |
| Bangladesh | -0.037 | -0.061 | -0.032 | -0.042 | -0.028 | -0.035 |
| Barbados | -0.003 | 0.000 | -0.001 | 0.000 | 0.088 | 0.000 |
| Belarus | 0.568 | 0.046 | 0.186 | 0.018 | 0.473 | 0.011 |
| Belgium | 0.318 | 0.108 | 0.137 | 0.052 | 0.093 | 0.035 |
| Belize | 0.000 | -0.001 | 0.000 | 0.000 | 0.000 | 0.000 |
| Benin | 0.047 | -0.012 | 0.009 | -0.014 | 0.003 | -0.013 |
| Bermuda | 0.000 | 0.000 | 0.000 | 0.000 | 0.000 | 0.000 |
| Bhutan | 0.000 | -0.001 | -0.001 | -0.001 | 0.000 | -0.001 |
| Bolivia | -0.004 | -0.008 | -0.004 | -0.006 | -0.004 | -0.005 |
| Bosnia and Herzegovina | 0.083 | 0.002 | 0.026 | 0.001 | 0.535 | 0.000 |
| Botswana | 0.011 | 0.004 | 0.005 | 0.002 | 0.003 | 0.001 |
| Brazil | 0.249 | 0.068 | 0.093 | 0.012 | 0.050 | -0.003 |
| Brunei | 0.002 | 0.001 | 0.001 | 0.000 | 0.001 | 0.000 |
| Bulgaria | 0.132 | 0.002 | 0.036 | 0.000 | 0.005 | 0.000 |
| Burkina Faso | 0.089 | -0.005 | 0.025 | -0.010 | 0.038 | -0.011 |

| Country | | | | | | |
|---|---|---|---|---|---|---|
| Burundi | 0.020 | -0.007 | 0.003 | -0.007 | 0.006 | -0.006 |
| Coted'Ivoire | 0.083 | -0.031 | 0.019 | -0.029 | 0.005 | -0.027 |
| Cambodia | -0.009 | -0.022 | -0.011 | -0.016 | -0.010 | -0.014 |
| Cameroon | 0.336 | -0.020 | 0.077 | -0.022 | 0.069 | -0.021 |
| Canada | 0.561 | 0.177 | 0.233 | 0.072 | 0.147 | 0.042 |
| Cape Verde | 0.001 | 0.000 | 0.001 | 0.000 | 0.000 | 0.000 |
| Central African Republic | 0.003 | -0.003 | 0.000 | -0.002 | 0.000 | -0.002 |
| Chad | 0.010 | -0.014 | -0.003 | -0.012 | -0.004 | -0.011 |
| Chile | 0.000 | -0.017 | -0.004 | -0.012 | -0.004 | -0.009 |
| China | 8.096 | 2.408 | 3.705 | 0.800 | 2.481 | 0.354 |
| Colombia | -0.001 | -0.018 | -0.007 | -0.015 | -0.008 | -0.014 |
| Comoros | 0.012 | -0.001 | 0.003 | -0.001 | 0.005 | -0.001 |
| Congo [DRC] | 0.096 | 0.002 | 0.030 | -0.010 | 0.017 | -0.012 |
| Congo [Republic] | 0.068 | -0.033 | 0.012 | -0.023 | -0.132 | -0.019 |
| Costa Rica | 0.005 | -0.007 | 0.000 | -0.005 | -0.001 | -0.004 |
| Croatia | 0.095 | 0.007 | 0.031 | 0.003 | 0.010 | 0.002 |
| Cuba | 0.004 | -0.010 | -0.001 | -0.007 | -0.002 | -0.006 |
| Cyprus | 0.005 | 0.001 | 0.002 | 0.000 | 0.001 | 0.000 |
| Czech Republic | 0.214 | 0.018 | -0.001 | 0.004 | 0.056 | 0.000 |
| Denmark | 0.129 | 0.027 | 0.050 | 0.008 | 0.031 | 0.003 |
| Djibouti | 0.004 | 0.002 | 0.002 | 0.001 | 0.001 | 0.001 |
| Dominica | 0.000 | 0.000 | 0.000 | 0.000 | 0.000 | 0.000 |
| Dominican Republic | 0.017 | -0.003 | 0.005 | -0.003 | 0.002 | -0.003 |
| Ecuador | -0.002 | -0.010 | -0.005 | -0.008 | -0.005 | -0.007 |
| Egypt | 0.729 | 0.151 | 0.282 | 0.042 | 0.198 | 0.010 |
| El Salvador | 0.000 | -0.003 | -0.001 | -0.002 | -0.001 | -0.002 |
| Equatorial Guinea | 0.004 | -0.011 | -0.004 | -0.009 | -0.005 | -0.008 |
| Eritrea | 0.003 | 0.000 | 0.001 | 0.000 | 0.001 | 0.000 |
| Estonia | 0.004 | 0.018 | 0.027 | 0.008 | 0.105 | 0.005 |
| Ethiopia | 0.315 | 0.000 | 0.092 | -0.026 | 0.131 | -0.029 |
| Faroe Islands | 0.000 | 0.000 | 0.000 | 0.000 | 0.000 | 0.000 |
| Fiji | 0.001 | -0.001 | 0.000 | -0.001 | 0.000 | -0.001 |
| Finland | 0.257 | 0.054 | 0.096 | 0.024 | 0.083 | 0.015 |
| France | 1.413 | 0.376 | 0.742 | 0.165 | 0.411 | 0.102 |
| Gabon | 0.018 | 0.004 | 0.007 | 0.001 | 0.004 | 0.000 |
| Gambia | 0.021 | 0.000 | 0.006 | -0.001 | 0.013 | -0.001 |
| Georgia | 0.167 | 0.004 | 0.047 | 0.000 | -0.429 | -0.001 |
| Germany | 2.340 | 0.736 | 1.133 | 0.356 | 0.710 | 0.241 |
| Ghana | 0.401 | 0.003 | 0.178 | -0.020 | -0.405 | -0.023 |
| Greece | 0.006 | -0.014 | -0.003 | -0.011 | -0.004 | -0.010 |
| Greenland | 0.000 | 0.000 | 0.000 | 0.000 | 0.000 | 0.000 |
| Grenada | 0.000 | 0.000 | 0.000 | 0.000 | 0.000 | 0.000 |
| Guatemala | -0.017 | -0.042 | -0.017 | -0.027 | -0.015 | -0.022 |
| Guinea | 0.175 | 0.000 | 0.037 | -0.003 | 0.127 | -0.004 |
| Guinea-Bissau | 0.020 | -0.001 | 0.005 | -0.002 | 0.015 | -0.002 |
| Guyana | 0.000 | -0.001 | 0.000 | -0.001 | 0.000 | 0.000 |
| Haiti | 0.030 | -0.009 | 0.007 | -0.006 | -0.058 | -0.005 |
| Honduras | 0.000 | -0.008 | -0.002 | -0.005 | -0.002 | -0.004 |
| Hong Kong | 0.053 | 0.029 | 0.028 | 0.016 | 0.019 | 0.012 |
| Hungary | 0.063 | 0.029 | 0.025 | 0.012 | 0.016 | 0.007 |
| Iceland | 0.002 | -0.001 | 0.000 | -0.001 | 0.000 | -0.001 |
| India | -0.960 | -1.105 | -0.686 | -0.738 | -0.564 | -0.596 |
| Indonesia | -0.066 | -0.151 | -0.101 | -0.126 | -0.102 | -0.110 |
| Iran | 0.679 | 0.000 | 0.815 | -0.033 | 0.545 | -0.037 |
| Iraq | 0.274 | -0.017 | 0.221 | -0.033 | 0.130 | -0.034 |
| Ireland | 0.070 | 0.000 | 0.022 | -0.006 | 0.012 | -0.006 |
| Isle of Man | 0.000 | 0.000 | 0.000 | 0.000 | 0.000 | 0.000 |
| Israel | 0.087 | -0.043 | 0.024 | -0.030 | -0.045 | -0.024 |
| Italy | 0.680 | 0.226 | 0.284 | 0.098 | 0.189 | 0.060 |
| Jamaica | 0.002 | -0.001 | 0.000 | 0.000 | 0.000 | 0.000 |
| Japan | 0.237 | 0.103 | 0.130 | 0.071 | 0.096 | 0.058 |
| Jordan | 0.012 | 0.001 | 0.005 | 0.000 | 0.002 | 0.000 |
| Kazakhstan | 0.696 | 0.262 | 0.449 | 0.110 | 0.344 | 0.065 |
| Kenya | 0.261 | -0.014 | 0.067 | -0.021 | 0.068 | -0.021 |
| Kiribati | 0.000 | 0.000 | 0.000 | 0.000 | 0.000 | 0.000 |
| Kosovo | 0.000 | 0.000 | 0.000 | 0.000 | 0.000 | 0.000 |
| Kuwait | 0.077 | 0.039 | 0.041 | 0.024 | 0.029 | 0.018 |
| Kyrgyzstan | 0.148 | 0.005 | 0.046 | 0.000 | 0.086 | -0.001 |
| Laos | -0.012 | -0.020 | -0.010 | -0.014 | -0.009 | -0.011 |
| Latvia | 0.077 | 0.015 | 0.036 | 0.006 | -0.097 | 0.004 |
| Lebanon | 0.006 | -0.005 | 0.001 | -0.004 | 0.000 | -0.003 |
| Lesotho | 0.001 | 0.000 | 0.000 | 0.000 | 0.000 | 0.000 |
| Liberia | 0.005 | -0.001 | 0.001 | -0.001 | 0.001 | -0.001 |
| Libya | 0.084 | 0.006 | 0.027 | -0.001 | 0.036 | -0.002 |
| Liechtenstein | 0.002 | 0.000 | 0.000 | 0.000 | 0.011 | 0.000 |
| Lithuania | -0.013 | 0.018 | 0.041 | 0.008 | 0.074 | 0.005 |
| Luxembourg | 0.055 | 0.023 | 0.025 | 0.011 | 0.017 | 0.008 |
| Macau | 0.002 | -0.005 | 0.000 | -0.003 | -0.001 | -0.003 |
| Macedonia [FYROM] | 0.003 | 0.001 | 0.001 | 0.000 | 0.001 | 0.000 |
| Madagascar | 0.023 | -0.026 | -0.003 | -0.021 | 0.000 | -0.019 |
| Malawi | 0.130 | 0.030 | 0.049 | 0.011 | 0.047 | 0.006 |
| Malaysia | -0.001 | -0.028 | -0.016 | -0.024 | -0.018 | -0.022 |
| Maldives | 0.005 | 0.000 | 0.001 | 0.000 | 0.001 | 0.000 |
| Mali | 0.209 | -0.014 | 0.061 | -0.019 | 0.165 | -0.019 |
| Malta | 0.005 | 0.001 | 0.002 | 0.000 | 0.000 | 0.000 |
| Marshall Islands | 0.000 | 0.000 | 0.000 | 0.000 | 0.000 | 0.000 |
| Mauritania | 0.147 | 0.004 | 0.039 | 0.000 | 0.144 | -0.001 |
| Mauritius | 0.003 | 0.000 | 0.001 | 0.000 | 0.000 | 0.000 |
| Mexico | -0.150 | -0.205 | -0.106 | -0.131 | -0.087 | -0.104 |
| Micronesia | 0.000 | 0.000 | 0.000 | 0.000 | 0.000 | 0.000 |
| Moldova | 0.049 | 0.004 | 0.021 | 0.001 | -0.335 | 0.000 |
| Monaco | 0.000 | 0.000 | 0.000 | 0.000 | 0.000 | 0.000 |
| Mongolia | 0.493 | 0.012 | 0.154 | 0.004 | -0.591 | 0.002 |
| Montenegro | 0.000 | 0.000 | 0.000 | 0.000 | 0.000 | 0.000 |
| Morocco | 0.618 | 0.085 | 0.322 | 0.036 | 0.049 | 0.021 |
| Mozambique | 0.032 | -0.022 | 0.002 | -0.020 | -0.004 | -0.018 |
| Namibia | 0.012 | 0.002 | 0.004 | 0.000 | 0.002 | 0.000 |
| Nepal | -0.015 | -0.020 | -0.011 | -0.013 | -0.010 | -0.011 |
| Netherlands | 0.572 | 0.118 | 0.214 | 0.049 | 0.196 | 0.029 |
| New Zealand | -0.001 | -0.019 | -0.004 | -0.012 | -0.004 | -0.009 |
| Nicaragua | 0.001 | -0.007 | -0.001 | -0.005 | -0.002 | -0.004 |
| Niger | 0.154 | -0.031 | 0.036 | -0.033 | 0.080 | -0.031 |
| Nigeria | 2.763 | 0.352 | 1.142 | 0.009 | 0.675 | -0.072 |
| Norway | 0.239 | 0.087 | -0.092 | 0.041 | 0.069 | 0.027 |
| Oman | 0.127 | -0.004 | 0.038 | -0.005 | 0.037 | -0.005 |
| Pakistan | 0.029 | -0.133 | -0.033 | -0.099 | -0.043 | -0.083 |
| Palau | 0.000 | 0.000 | 0.000 | 0.000 | 0.000 | 0.000 |
| Palestinian Territories | 0.000 | 0.000 | 0.000 | 0.000 | 0.000 | 0.000 |
| Panama | 0.012 | 0.000 | 0.004 | -0.001 | 0.002 | -0.001 |
| Papua New Guinea | -0.005 | -0.009 | -0.007 | -0.007 | -0.007 | -0.006 |
| Paraguay | -0.012 | -0.012 | -0.008 | -0.008 | -0.006 | -0.007 |
| Peru | 0.010 | 0.013 | 0.002 | 0.002 | 0.000 | -0.001 |
| Philippines | -0.022 | -0.078 | -0.045 | -0.065 | -0.050 | -0.057 |
| Poland | 0.165 | 0.065 | 0.104 | 0.027 | 0.074 | 0.016 |
| Portugal | 0.068 | 0.018 | 0.027 | 0.007 | 0.018 | 0.004 |
| Puerto Rico | 0.008 | 0.001 | 0.004 | 0.001 | 0.002 | 0.000 |
| Qatar | 0.399 | 0.045 | 0.133 | 0.007 | 0.109 | -0.001 |
| Romania | 0.039 | 0.017 | 0.136 | 0.000 | 0.078 | -0.004 |
| Russia | 3.109 | 1.221 | 1.646 | 0.532 | 1.189 | 0.329 |
| Rwanda | 0.044 | -0.008 | 0.011 | -0.009 | 0.023 | -0.008 |
| São Tomé and Príncipe | 0.000 | 0.000 | 0.000 | 0.000 | 0.000 | 0.000 |
| Saint Kitts and Nevis | 0.000 | 0.000 | 0.000 | 0.000 | 0.000 | 0.000 |
| Saint Lucia | 0.000 | 0.000 | 0.000 | 0.000 | 0.000 | 0.000 |
| Saint Vincent and the Grenadines | 0.000 | 0.000 | 0.000 | 0.000 | 0.000 | 0.000 |
| Samoa | 0.000 | 0.000 | 0.000 | 0.000 | 0.000 | 0.000 |
| San Marino | 0.000 | 0.000 | 0.000 | 0.000 | 0.000 | 0.000 |
| Saudi Arabia | 0.301 | 0.036 | 0.121 | 0.015 | 0.084 | 0.009 |
| Senegal | 0.107 | 0.013 | 0.040 | 0.002 | 0.028 | -0.001 |
| Serbia | 0.000 | 0.000 | 0.000 | 0.000 | 0.000 | 0.000 |
| Seychelles | 0.000 | 0.000 | 0.000 | 0.000 | 0.000 | 0.000 |
| Sierra Leone | 0.025 | -0.004 | 0.006 | -0.005 | 0.011 | -0.004 |
| Singapore | 0.055 | 0.038 | 0.032 | 0.022 | 0.024 | 0.016 |
| Slovakia | 0.031 | 0.008 | 0.012 | 0.003 | 0.008 | 0.002 |
| Slovenia | 0.025 | 0.007 | 0.010 | 0.003 | 0.007 | 0.002 |
| Solomon Islands | 0.001 | 0.000 | 0.000 | 0.000 | 0.000 | 0.000 |
| South Africa | 0.157 | 0.048 | 0.067 | 0.020 | 0.043 | 0.012 |
| South Korea | 0.061 | 0.010 | 0.022 | 0.008 | 0.012 | 0.007 |
| Spain | 0.546 | 0.167 | 0.223 | 0.073 | 0.153 | 0.045 |
| Sri Lanka | -0.003 | -0.011 | -0.005 | -0.008 | -0.005 | -0.007 |
| Sudan | 0.092 | 0.013 | 0.031 | -0.002 | 0.014 | -0.007 |
| Suriname | -0.001 | -0.001 | -0.001 | -0.001 | 0.000 | -0.001 |
| Swaziland | 0.001 | 0.000 | 0.000 | 0.000 | 0.000 | 0.000 |
| Sweden | 0.493 | 0.134 | 0.202 | 0.058 | 0.136 | 0.036 |
| Switzerland | 0.449 | 0.133 | 0.187 | 0.058 | 0.125 | 0.036 |

| Country | | | | | | |
|---|---|---|---|---|---|---|
| Syria | 0.030 | -0.009 | 0.009 | -0.007 | 0.001 | -0.006 |
| Tajikistan | 0.186 | -0.001 | 0.051 | -0.005 | 0.129 | -0.006 |
| Tanzania | 0.246 | -0.157 | 0.011 | -0.134 | 0.150 | -0.118 |
| Thailand | 0.026 | -0.004 | 0.000 | -0.009 | -0.004 | -0.010 |
| Timor-Leste | -0.003 | -0.004 | -0.002 | -0.002 | -0.002 | -0.002 |
| Togo | 0.028 | -0.005 | 0.007 | -0.005 | 0.018 | -0.005 |
| Tonga | 0.001 | 0.000 | 0.000 | 0.000 | 0.028 | 0.000 |
| Trinidad and Tobago | 0.003 | 0.001 | 0.001 | 0.001 | 0.001 | 0.001 |
| Tunisia | 0.216 | 0.027 | 0.080 | 0.012 | -0.121 | 0.007 |
| Turkey | 0.389 | -0.122 | 0.101 | -0.088 | 0.052 | -0.073 |
| Turkmenistan | 0.401 | 0.036 | 0.281 | 0.011 | 0.104 | 0.004 |
| Tuvalu | 0.000 | 0.000 | 0.000 | 0.000 | 0.000 | 0.000 |
| Uganda | 0.433 | -0.035 | 0.110 | -0.041 | 0.258 | -0.039 |
| Ukraine | 0.224 | 0.038 | 0.089 | 0.015 | -0.221 | 0.008 |
| United Arab Emirates | 1.121 | 0.066 | 0.394 | 0.016 | 0.304 | 0.003 |
| United Kingdom | 1.211 | 0.345 | 0.507 | 0.152 | 0.330 | 0.093 |
| United States | 5.773 | 3.229 | 2.675 | 1.834 | 1.788 | 1.344 |
| Uruguay | 0.000 | -0.002 | 0.000 | -0.001 | 0.000 | -0.001 |
| Uzbekistan | 1.245 | 0.056 | 0.530 | 0.017 | -0.054 | 0.006 |
| Vanuatu | 0.002 | -0.001 | 0.001 | -0.001 | -0.133 | 0.000 |
| Venezuela | 0.009 | -0.004 | 0.001 | -0.006 | -0.001 | -0.006 |
| Vietnam | 0.010 | -0.035 | -0.020 | -0.034 | -0.023 | -0.031 |
| Yemen | 0.281 | -0.007 | 0.087 | -0.009 | 0.026 | -0.009 |
| Zambia | 0.108 | 0.010 | 0.037 | 0.000 | 0.033 | -0.002 |
| Zimbabwe | 0.017 | 0.001 | 0.006 | -0.001 | 0.003 | -0.002 |